\documentstyle[12pt,preprint,aps,psfig]{revtex}
\def\be{\begin{equation}}
\def\ee{\end{equation}}
\def\bea{\begin{eqnarray}}
\def\eea{\end{eqnarray}}
\def\d{\partial}
\def\lbd{\lambda}

\def\Gam{\Gamma}
\begin{document}
\baselineskip=16pt
\begin{title}
\hfill \flushright{\date{today}}
\vskip 2cm
\begin{center}
{\Large\bf  Derivative Corrections To The Sphaleron Energy }
\vskip 1.6cm
{\bf{Harnish Nagar and  Urjit A. Yajnik}} \\
\vskip .5cm
\it{Department of Physics, Indian Institute of Technology, Bombay,
Mumbai 400076, India}
\vskip .2cm
\vskip .3cm
\end{center}
\vskip .5cm
\begin{quotation}
\vskip .2cm

We investigate the effects of inhomogeneous scalar field
configurations on effective action at finite
temperature. Recently, Walliser {\it{et al.}} have calculated the leading
perturbative corrections to the wave function correction
term $Z(\phi,T)$, for the standard model at finite temperature.
Here we follow their results and apply them to calculate
the effective energy of the sphaleron solution, including
derivative correction term $Z(\phi,T)$.
The dominant contribution to $Z(\phi,T)$ comes from vector
boson loops. We also study the possible effects of corrected
sphaleron energy on the baryon number violation rate in the
standard model.
\end{quotation}
\end{title}
\section{Introduction}
Baryon number violation processes in the electroweak 
theory play an important role in the understanding of baryon asymmetry
of the universe. It is known that anomaly induced baryon number 
violation rate at zero  temperature \cite{hooft} is considerably
suppressed. \par
At high temperatures the rate formula is derived by 
Affleck-Arnold mechanism \cite{affleck}, \cite{mac}. At electroweak
symmetry breaking temperatures sphaleron configuration \cite{manton} separates
different Chern-Simons vacua. Each time system makes a transition
through sphaleron configuration, anomaly induced baryon number 
violation occurs. The sphaleron rate \cite{shap} is given by,
\be
 \Delta \propto ( \beta E_{sph})^{8} \exp [- \beta E_{sph} ]
\label{e50}\ee \noindent 
Where $\beta=T^{-1}$ and $E_{sph}$ is the energy of the sphaleron solution.
\par
We have reanalyzed this rate formula and found that for all values
of Higgs boson mass sphaleron rate is unsuppressed and any net
baryon excess generated prior to electroweak phase transition
is washed out. This might indicate  that present baryon asymmetry can not be
explained in the framework of minimal standard model. \par
In this article we review derivative corrections to the effective action 
and apply it to calculate corrections to the sphaleron energy.
We find that correction to sphaleron energy is negative and sphaleron rate
is strongly unsuppressed. This again indicates complete erasure of
baryon asymmetry of the universe in the minimal standard model.
\section{The Derivative Correction Term}
Effective action can be written as an expansion in terms of
fields and its derivatives (momentum space expansion),
\be
\Gam[\phi(x)] = \int d^4x \left[ \frac{1}{2} \{ 1+Z(\phi) \} (\d^{\mu}\phi)^{\dagger} (\d_{\mu}\phi)
- V(\phi) + higher \ order \ derivative \ terms\ .....\right]
\ee \noindent \par 
Here $Z(\phi)$ is the derivative (wave function) correction term. 
We will confine to kinetic energy correction term and hence neglect
higher order derivative terms.
We point out that effective potential can be extracted out  
by expanding effective action in a constant background field $\bar{\phi}$.
In Euclidean space finite temperature effective action is written as,
\be
\Gamma[\phi,T] = \beta \int d^3x \left[ \frac{1}{2} 
\{ (1 + Z(\phi,T) \} (\d^{\mu}\phi)^{\dagger} (\d_{\mu}\phi)
+ V(\phi,T) \right]
\ee \noindent \par
Where $\beta= \frac{1}{T}$ and we have considered time independent fields, 
$\phi(x)= \phi(\vec{x})$.
In the following we briefly describe effects of Z($\phi$,T) term for a 
bubble nucleation  theory \cite{linde}, \cite{jung}.
\par
The free energy of the bubble solution is of the form,
\be
F[\phi,T] = \int d^3x \left[ \frac{1}{2} |\vec{\nabla}\phi|^2 + V(\phi,T) \right]
\label{e1}\ee \noindent \par
The exact expression for the free energy of the bubble solution is given by
F[$\bar{\phi}$(r),T], with $\bar\phi$(r), r=$|\vec{x}|$ representing bubble solution.
V($\phi$,T) is temperature dependent effective potential,
$$ V(\phi,T) = \frac{1}{2} D T^2 \phi^2 - ET \phi^3 + \frac{\lbd \phi^4 }
{4} $$
Where constants D and E depend only on the scalar self coupling constant $\lambda$. 
V($\phi$,T) is shown in Fig. 1.
We note that V($\phi$,T) describes first order phase transition with two
degenerate minima at critical temperature $ T=T_{c},\  at \ T <  T_{c}$ there
is one local minimum at $\phi=0$ and a global minimum at $\phi=\phi_{min}$. 
\par
The effect of $Z(\phi,T) \neq 0 $ term can be understood by considering
a scalar field transformation in equation (\ref{e1}),
$$ \tilde{\phi} (r)  =  \int d \phi\sqrt{1 + Z( \phi,T)} $$
Now for Z $(\phi$,T) $>$ 0,  
$$ \frac{\d \tilde{\phi}(r)}{\d \phi(r)}= \sqrt{1 + Z(\phi,T)}
 > 0 $$
\noindent 
\newpage
\begin{center}
\begin{minipage}[ht]{7.75cm}
\hbox{\centerline{\psfig{figure=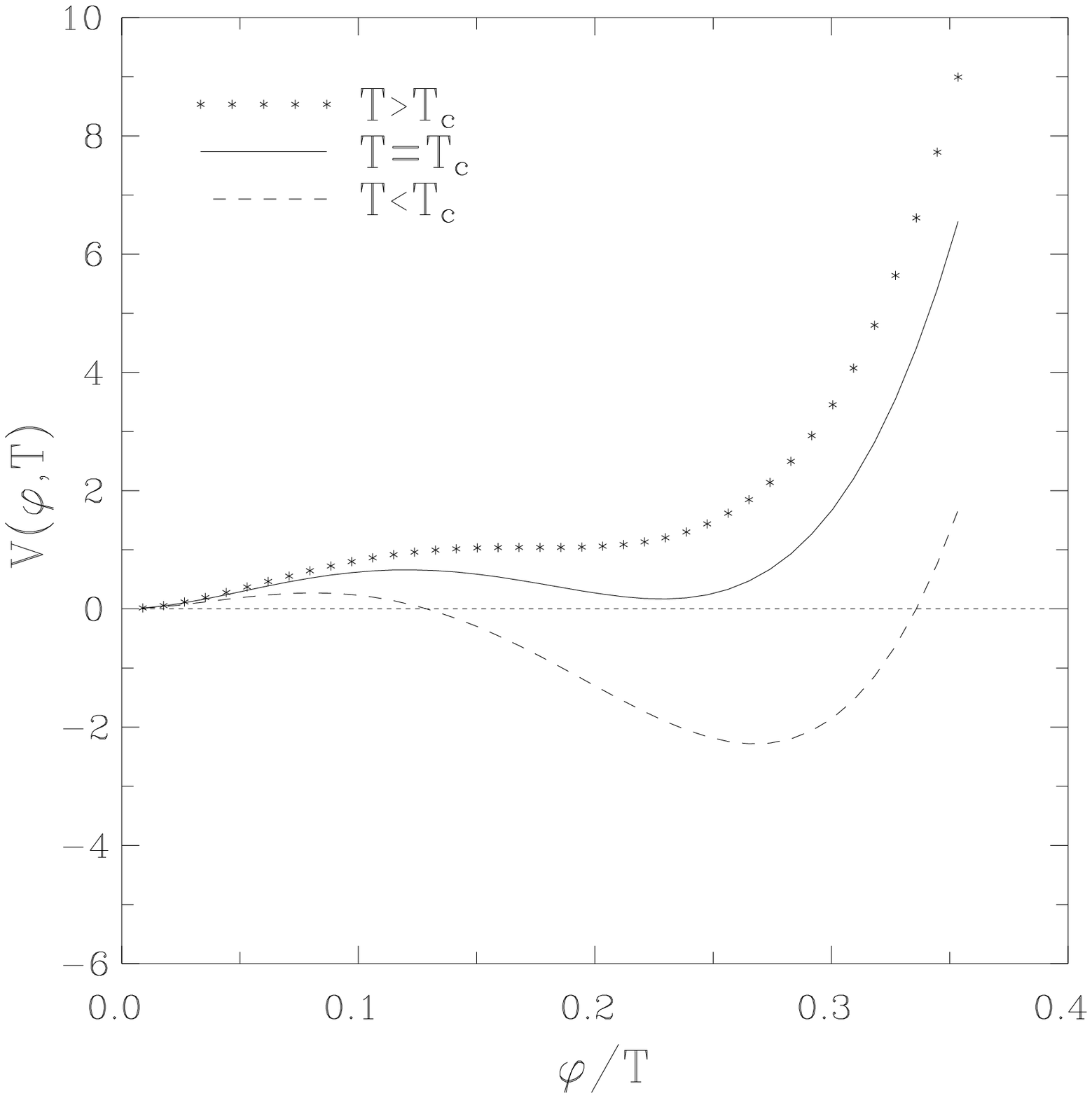,width=12cm}}}
\begin{center} 
Fig. 1. Shows effective potential at finite 
temperature. $T_c$ is the critical temperature. 
\end{center}
\end{minipage}
\hfil
\end{center}
\vskip .5cm
\par
This means that free energy $ F(\tilde{\phi},T)$ and V($\tilde{\phi}$,T) have
the same form and V($\tilde{\phi}$,T) is locally rescaled.
However the surface energy of the bubble changes significantly which in turn
affects free energy, $ F({\tilde{\bar\phi}},T)$. 
A positive value of $Z(\phi,T)$ increases the free energy and 
negative value of Z decreases the free energy of bubble solution. \par
Another important consequence of positive Z is that it enhances the
strength of phase transition. Since positive Z increases surface energy, to
counter effect this enhancement universe would have to
supercool further to complete the phase transition. This in turn
reduces the end temperature of phase transition $T_{e}$, thereby
increasing the strength of phase transition, $ \frac{ \phi_{min}}{T_e}$.
\section{Derivative Corrections to Sphaleron Energy}
Sphaleron  solution \cite{manton} is described by an SU(2) invariant 
time independent Higgs field theory,
without fermions.
The energy of the sphaleron solution is given by,
\be
E_{sph} = \int d^3x \left[ \frac{1}{4} {W_{ij}}^{a} {W_{ij}}^{a} +
 (D_i\phi)^{\dagger} (D_{i}\phi) + V(\phi) \right]
\ee
Where $W_{ij}^{a}$ is SU(2) gauge field tensor for gauge fields
$W_i^{a}(\vec{x})$ and $\phi(\vec{x})$ is Higgs scalar field.
$$ (D_{i}\phi)= \d_i \phi - \frac{1}{2} ig \sigma^a W_{i}^a \phi $$
$$ V(\phi,T=0) = \lambda (\phi^\dagger \phi - \frac{1}{2} v^2)^2 $$
\noindent \par
g is SU(2) gauge coupling constant.
$\lambda$ and v represents scalar self coupling constant and zero temperature
vacuum expectation value respectively. We know that in the standard model,
tree level Higgs boson mass is related to the scalar coupling constant by,
$$ {M_h}^2 = 2 \lambda v^2 $$
With a derivative correction term we write the effective energy at
finite temperatures as,
\be
E_{sph} = \int d^3x \left[ \frac{1}{4} {W_{ij}}^{a} {W_{ij}}^{a} +
\{1 + Z_{SM}(\phi,T)\}(D_i\phi)^{\dagger} 
(D_{i}\phi) + V(\phi,T) \right]
\label{e6}\ee
Where $Z_{SM}$ is the standard model contribution to wave function correction
term, it can be written as,
$$ Z_{SM} = Z_{scalar} + Z_{vector} + Z_{fermi} $$ \par
Recently complete calculations for $Z_{SM}$ have been performed 
by  Jungnickel {\it{et al.}} \cite{jung}.
Their finding is that scalar boson and fermion contributions are considerably
suppressed compare to vector boson contributions. At high temperatures,
scalar and vector boson contributions at T$\neq$0 dominate 
over their respective contributions at T=0.
Hence we need only write,
\be
Z_{scalar}^{T\neq0}(\phi,T) = \frac{\lbd^2 T \phi^2}{16 \pi}
\left( \frac{3}{M_{\phi}^3} +  \frac{9\pi_{\phi}}{2M_{\phi}^5}
+ \frac{1}{M_{\chi}^3} + \frac{3 \pi_{\chi}}{2 M_{\chi}^5}
\right)  \ee
\noindent \par
Where $M_{\phi}$ and $M_{\chi}$ are mass matrices for Higgs and
Goldstone boson and $\pi_{\phi}$ and $\pi_{\chi}$ are corresponding
self energy corrections.
\be
 Z_{vector}^{T\neq0}(\phi,T) = - \frac{3 g^2 T}{4 \pi} \left(
\frac{m_w^4}{32 M_{L}^5} - \frac{5 m_w^2}{96 M_{L}^3} 
+  \frac{5 m_w^4}{16 M_{T}^5} -  \frac{41 m_w^2}{48 M_{T}^3}   
+ \frac{1}{M_{T}} \right)
\label{e31}\ee \noindent \par
Where $m_w$ is zero temperature mass of gauge boson and
$M_{L} \ and \ M_T $ are transverse and longitudinal masses for gauge
boson.
$$ M_{L}^2 =   {m_{w}}^2(\phi) + \pi_{L}(0) $$
$$   M_{T}^2 =   {m_{w}}^2(\phi) + \pi_{T}(0) $$
with,  $ {m_{w}}^2(\phi)= \frac{1}{2} g^2 {\phi}^2 $. 
g is SU(2) gauge coupling constant and $\pi_{L}(0)$ 
,$ \pi_T(0)$ are self
energy contributions to gauge boson masses. \par
The full fermionic contribution is given by,
\be
Z_{fermi}(\phi,T)= \frac{f_{t}^2}{96 \pi^2}
\left(7 + 12 \gamma_E + 12 \ln \frac{m_{ren}}{\pi T} \right)
\ee \noindent 
\newpage
\begin{center}
\begin{tabular}{||c|c|c|c||} \hline 
$\frac{\lambda}{g^2}$  & $[E_{sph}$ (Z=0)] (5TeV) & Temperature(GeV) & Energy correction(5TeV) \\ \hline  
.001     & 1.606    & 100   & -0.080 \\ \cline{3-4}  
          &           & 246   & -.20 \\ \hline
.01      & 1.675     & 100  &  -0.076 \\  \cline{3-4}
          &           & 246   & -.19 \\ \hline
.10      & 1.83     &  100  &  -0.071 \\  \cline{3-4}
          &           & 246   & -.177 \\ \hline
1.0      & 2.099    &  100  &  -0.067  \\  \cline{3-4}
          &           & 246   & -.167 \\ \hline
\end{tabular}
\end{center}
\begin{center}
Table [1] Shows sphaleron energy corrections for different values of
scalar coupling constant. 
\end{center}
\vskip .5cm
\par
Where $f_t$ is top quark Yukawa coupling to Higgs boson and $m_{ren}$ is
renormalized mass matrix for top quark. $\gamma_E \simeq .577$. For details 
we refer article by  Jungnickel {\it{et al.}} \cite{jung}. \par
We had already stressed the fact that vector boson contribution to $Z(\phi,T)$
dominates over scalar and fermion contributions hence we need only include vector
boson corrections, that is,
                    $$Z_{SM}=Z_{vector}$$
\section{Results}
We have numerically evaluated the expression (\ref{e6}). 
Table [1] shows sphaleron energy for different values of 
scalar coupling constant. The correction is 
negative and for temperature range 80 to 120 GeV, the
numerical value is very small. Energy correction is plotted with
temperature for different values of $\frac{\lbd}{g^2}$ in Fig. 2.
We mention that in the range we  have considered for $\frac{\lambda}{g^2}$ (.001-1.0), the
Higgs boson mass varies from 10 GeV to 225 GeV. For this Higgs mass
the critical temperature for electroweak phase transition varies from
50 GeV to 270 GeV.
\par
We find that numerical value of correction is very small, however
the effect is to decrease the total energy of the sphaleron.
We have calculated the sphaleron rate, equation (\ref{e50}), with corrected
sphaleron energy and found that sphaleron rate is strongly 
unsuppressed. 
\newpage
\begin{center}
\begin{minipage}[ht]{12cm}
\hbox{\centerline{\psfig{figure=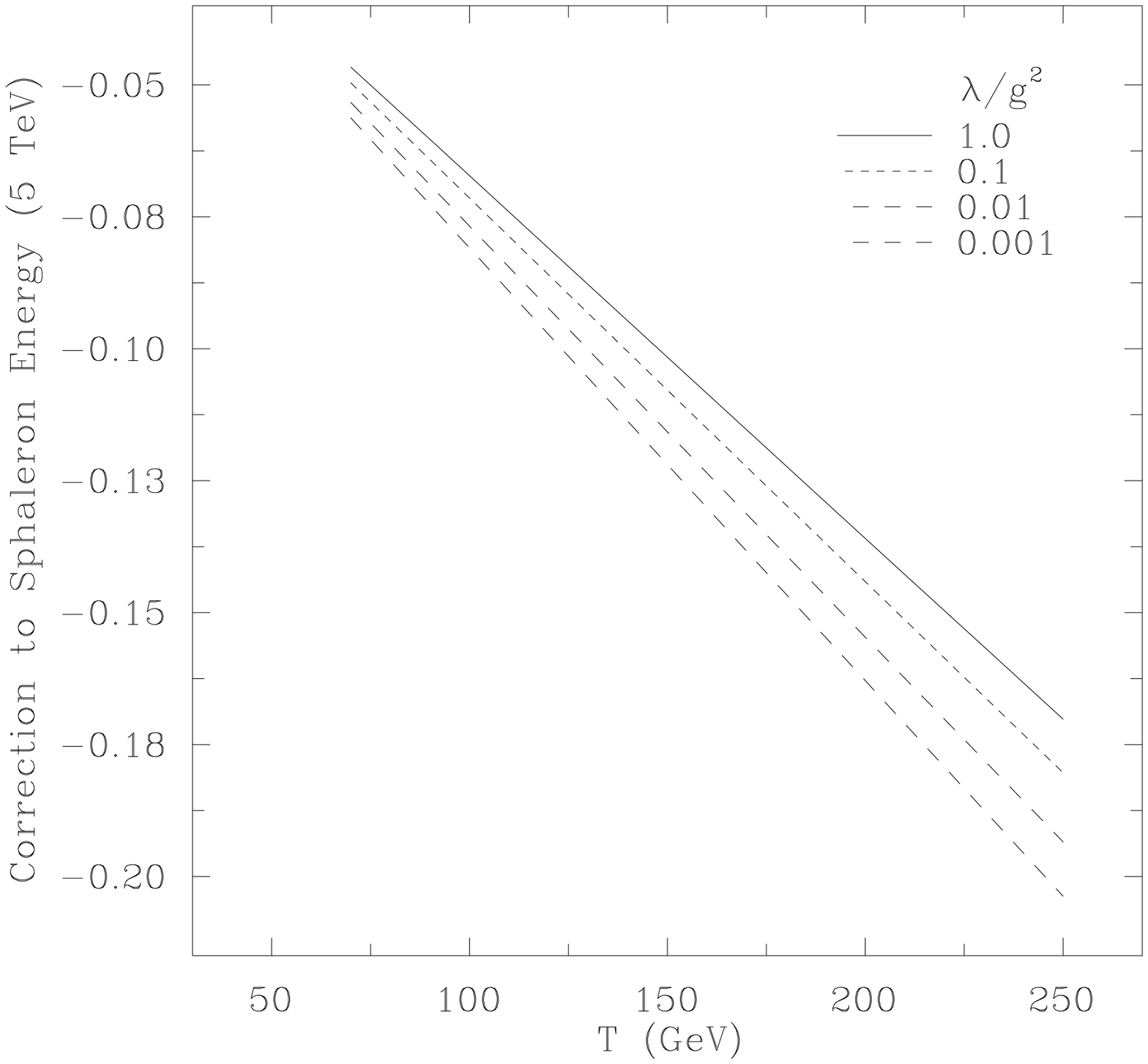,width=12cm}}}
\begin{center} 
Fig. 2. Shows correction to sphaleron energy at different
temperatures for different values of scalar coupling constant $\lambda$.
\end{center}
\end{minipage}
\end{center}
\vskip .5cm
\par Hence our results only support the current status of electroweak 
baryogenesis that is, present baryon asymmetry of the 
universe can not be explained in the framework of minimal standard model. 
We have considered one loop effective action including self energy
corrections to calculate derivative correction term $Z(\phi,T)$, it 
remains to check the effect of
two loop corrections on $Z(\phi,T)$ and hence on sphaleron energy. \\
{\bf Acknowledgments} \\
This work partially supported by Department of Science and Technology, India.
\bibliographystyle{unsrt}

\end{document}